# Combinatorial logic devices based on a multi-path active ring circuit

Alexander Khitun and Michael Balinskiy

Electrical Engineering Department, University of California - Riverside, Riverside, CA, USA, 92521

**Abstract.** In this work, we describe a logic device in which an act of computation is associated with finding a path connecting input and output ports. The device is based on an active ring circuit comprising electric and magnetic parts. The electric part includes an amplifier, a phase shifter, and an attenuator. The magnetic part is a multi-port magnetic matrix comprising delay lines and frequency filters. Signals propagating on different paths may accumulate different phase shifts. Auto-oscillations occur in the circuit when the magnetic and electric parts match each other to meet the resonance amplitude and phase conditions. The system naturally searches for a resonance path that depends on the position of the electric phase shifter and amplification level. The path is detected by the set of power sensors. The proposed logic device can be used for solving a variety of computational problems. We present the results of numerical modeling illustrating prime factorization and finding the shortest path connected selected points on the mesh. We also present experimental data on the proof-of-the-concept experiment for the two-path device. The magnetic part consists of two waveguides made of single-crystal yttrium iron garnet $Y_3Fe_2(FeO_4)_3$ (YIG) films. Different phase shifts per delay line are achieved by adjusting the magnitude and direction of the bias magnetic field. The auto-oscillation signal changes the propagation path in the magnetic matrix depending on the position of the outer electric phase shifter. The power difference between the active and passive paths exceeds 40 dBm at room temperature. The described logic devices are robust, deterministic, and operate at room temperature. The number of possible paths increases factorial with the size of the mesh. It may be possible to encode information in paths and retrieve it using the external phase shifters and attenuators. Potentially, combinatorial logic devices may compete with quantum computers in functional throughput. Physical limits and constraints are also discussed.

**Introduction**

Modern computing devices consist of a large number of electric switches assembled in Boolean logic gates. It was George Bool, a great mathematician and philosopher of the XIX century, who came up with the idea of describing the process of thinking as a sequence of elementary logic operations. He proved that all processes can be accomplished with a set of logic gates operating with just two logic variables True of False or 0 and 1. In his book "The Laws of Thought", he described the basic logic gates such as NOT, OR, and AND [1]. Universality is the most appealing property of Boolean logic. According to the "Laws of Thought," *any* computational task can be decomposed into several elementary logic gates. Later on, in the XX century, Claude Shannon read the book of George Boole and came up with the idea of implementing Boolean logic gates with electric switches. Being a student at the Massachusetts Institute of Technology, Shannon described in his master thesis the implementation of Boolean logic gates with electromechanical relays [2]. In this approach, logic 0 and logic 1 in Boolean algebra became associated with the two levels of voltage (e.g., 0 V and 3 V). Logic functionality is defined by the configuration of switches in the circuit. The act of computation is related to the switching of the output voltage depending on the input(s). It was great progress since then in the development of computing devices by elaborating electric switches starting from electromechanical relays and vacuum tubes towards solid-state transistors [3]. Currently, complementary metal–oxide–semiconductor (CMOS) is the most elaborated electric switch enabling fast and energy-efficient electric current control [4]. The characteristic size of the silicon transistor (i.e. the channel length) has reached the sub 10 nm scale while the switching frequency is in the GHz range. Functional throughput is a commonly accepted metric for logic devices evaluation [5]. It can be calculated as follows:

$$Functional\ throughput = \frac{number\ of\ operations}{area \times time}. \qquad (1)$$

Boolean logic gates accomplish one operation at a time. The structure of transistor-based Boolean logic gates remains mainly unchanged for the last 50 years. The major enhancement in the functional throughput is associated with scaling down the size of transistors and increasing their switching frequency. The scaling down allows us to integrate a larger number of elements on the integrated circuit. According to Moore's law, the number of transistors per integrated circuit doubles every 18 months [6]. Current CMOS-based processors are extremely efficient in making basic mathematic operations such as addition and multiplication. At the same time, there are some specific types of problems (e.g., prime factorization, traveling salesman, the bridges of Konigsberg) that require a large number of operations for Boolean type logic gates.

The rapid development of the Internet of Things (IoT) and bioinformatics in the XXI century dictates a need for special type logic/memory devices for fast database search. These special task devices may be neither Boolean nor consist of electric switches. A quantum computer is an example of a non-Boolean logic device that is proven to be fundamentally more efficient than conventional logic devices in prime factorization and database search [7]. Quantum computer does not have to be of small size and/or fast switching. The ability to check a large number of input combinations in parallel is the main advantage of quantum computers. For example, a quantum computer with $n$ inputs may check in parallel $2^n$ possible input combinations. It would take $2^n$ consecutive operations for the conventional Boolean-type computer to solve the same problem.

In this work, we describe the operation of a logic device that can find one path out of many possible by exploiting the ability of an active ring circuit to self-adjust to a resonance frequency. Positive feedback spin-wave systems, often termed spin-wave active rings, are used in electronics as coherent microwave sources [8,9]. The basis for such an active ring is a dispersive spin-wave waveguide with exciting and receiving antennas connected via a variable-gain electrical feedback loop. If the correct gain and phase conditions are met, a monochromatic signal propagates in the ring and increases with time until nonlinear saturation takes place either in the spin-wave system or in the external amplifier [10]. We propose and describe a multi-path active ring circuit where signals propagating on different paths accumulate different phase shifts. Only certain signals propagating through resonant paths will be amplified while other paths will possess much lower energy flow.

**Results**

**Principle of operation**

To explain the principle of operation of the combinatorial logic device, we start with a spin-wave active ring circuit schematically shown in Fig.1(A). The circuit comprises electric and magnetic parts connected in series. The electric part includes a nonlinear amplifier G(p), variable phase shifter $\Psi$, and a controllable attenuator A. These parts are connected via standard coaxial cables. The magnetic part is a delay line - a waveguide made of material with low spin wave damping (e.g., YIG). Two microstrip antennas excite and receive spin waves through the waveguides. The detailed description of spin wave excitation and detection with microantennas can be found elsewhere [11]. Spin waves in propagating in the waveguide are much slower compared to the electromagnetic waves of the same frequency propagating in the coaxial cable. It provides a prominent phase shift Δ, which is, in general, frequency-dependent. Hereafter, we refer to the phase shifts accumulated inside the magnetic matrix as internal phase shifts. The phase shift provided by the electric phase shifter Ψ will be referred to as external phase shift. The signal circulating in the ring circuit exhibits a conversion from electromagnetic waves to spin waves and vice versa. The corresponding operator equation describing such a system has the form [12]:

$$L^{-1}\left(i\frac{d}{dt}\right)c(t) - G(p)c(t) = 0, \qquad (2)$$

where operator $L^{-1}\left(i\frac{d}{dt}\right)$ is a linear operator which describes the oscillating system with delay, $G(p)$ describes the nonlinear amplifier, the function $c(t)$ describes the complex amplitude of the auto-oscillation at the input, $p = |c(t)|^2$ is the signal power. In the frequency domain the operator L (*i d/dt*) can be described by a transfer function $L(\omega)$ that is defined as a Fourier transform of the impulse response function of the oscillating system and can be directly measured experimentally. The energy flow into the system is provided by the amplifier, while the passive oscillating system L determines the auto-oscillation frequency and provides positive damping.

The stable limit cycle of the auto-oscillator for the function $c(t)$ has the following form:

$$c_s(t) = \sqrt{p_s}e^{-i\phi(t)}, \qquad (3)$$

where $\phi(t) = \omega_s t + \phi_0$, $p_s$ and $\omega_s$ are the stationary free-running auto-oscillation power and frequency, respectively, and $\phi_0$ is an arbitrary initial phase of the auto-oscillation. The gain and phase conditions of the auto-oscillation are the following [12] :

$$\text{abs}[L(\omega_s)G(p_s)] \geq 1, \qquad (4.1)$$

$$\arg[L(\omega_s)] + \arg[G(p_s)] = 2\pi l, \qquad (4.2)$$

where *l = 1, 2, 3, …* is an integer number. The oscillations start when the gain provided by the amplifier G(p) exceeds the losses in the spin wave system and the sum of phases within the electric and spin wave parts matches the ring resonance condition. To simplify our consideration, we assume $\arg[L(\omega_s)] = \Delta$ , and $\arg[G(p_s)] = \Psi$ and neglect the effect of conducting cables on the phase shift accumulated by the signal.

The ability to self-adjust to the auto-oscillation frequency is the most appealing property of the active ring circuit. In case condition 4.1 is satisfied, the system is naturally searching for the frequency $\omega_s$ to satisfy condition 4.2. There is a sea of thermal magnons of different frequencies in the magnetic waveguide. Spin wave dispersion depends on several factors including material properties and the geometry of waveguide, the strength and the direction of the bias magnetic field. Spin waves of different frequencies may provide significantly different phase shifts to the propagating signal. However, only the signal on frequency(s) that met the phase condition 4.2 will be amplified. We want to emphasize the physical mechanism leading to the frequency self-adjustment. Signals of the same frequency circulating in the ring circuit superpose after each propagation cycle with a certain phase shift. Signals which met condition 4.2 superpose in-phase to each other which lead to the amplitude amplification. Other signals with some phase mismatch exhibit a sequence of constructive/destructive interference with no overall amplitude increase. There are results of numerical modeling shown in Fig.1(B) which are aimed to illustrate the amplification process. There are three curves of different colors corresponding to the three different cases. The black curve shows the results for the case $\Delta + \Psi = 2\pi$. The red and the blue curves correspond to $\Delta + \Psi = 2.1\pi$ and $\Delta + \Psi = 2.3\pi$, respectively. Only signal on the frequency satisfying the phase condition 4.2 (i.e., black curve) is amplified. They will come in phase round by round circulating in the ring circuit and amplified each time. The amplitude saturation appears either due to the physical limits of the non-linear amplifier or the magnetic saturation in the waveguide. The amplitudes of the other signals are much lower compared to the amplitude of the self-oscillation mode. Thus, the system starts with a superposition of a number of different frequencies and phases and only certain, resonant, frequencies will be amplified.

Next, let us consider an active ring circuit with two magnetic paths as shown in Fig.2(a). Each path includes a bandpass frequency filter, a phase shifter (delay line), and a power sensor. The filters are adjusted to different frequencies $f_1$ and $f_2$, respectively. The phase shifters are adjusted to two different phase shifts $\Delta_1$ and $\Delta_2$. The power sensors are aimed to detect the power flow through the path. More specifically, the power sensor detects whether the signal power exceeds the noise level or not. The modified circuit may transmit signals only on the two frequencies. There are three possible scenarios for signal propagation depending on the position of the electric phase shifter $\Psi$: (1) signal on frequency $f_1$ meets the self-oscillation conditions 4.1 and 4.2, the power goes through the upper path; (2) signal on frequency $f_2$ meets the self-oscillation conditions 4.1 and 4.2, the power goes through the lower path; (3) none of the two frequencies meet the resonance conditions, there are no self-oscillations in the circuit. In Fig.2(B), there are shown the results of numerical modeling on the amplitude of the self-oscillations depending on the position of the external phase shifter $\Psi$. As an example, we took $\Delta_1 = 1.7\pi$ and $\Delta_2 = \pi$. There are just two positions of the external phase shifter $\Psi$ at which circuit is in the self-oscillation. Importantly, the signals come through different paths. In the case of resonance on frequency $f_1$, most of the power comes through the upper path. Most of the power comes through the lower path if the resonance is on frequency $f_2$. The particular values of $f_1$ and $f_2$ are not important. *Signal propagation path is the computational output to be detected not the self-oscillation frequency.*

The two-path circuit can be cascaded as illustrated in Fig.3(a). It is shown a circuit consisting of the three two-path blocks. For simplicity, the power sensors are depicted as circles embedded in the delay lines. The green circle corresponds to the auto-oscillation, while the red circle means no energy flow above the noise level. Each of the bandpass filters has a set of four frequencies. For example, the bandpass filter on the first upper path transmits signals on frequencies $\{f_1, f_2, f_3, f_4\}$. The bandpass filter in the first block lower path transmits signals on frequencies $\{f_5, f_6, f_7, f_8\}$. The other bandpass filters have combinations of these eight frequencies. In this scenario, one input-output route is associated with one frequency. The delay lines are chosen to provide the following phase shifts. The phase shifts for the upper paths are $0.1\pi$, $0.3\pi$, and $1.5\pi$, respectively. The phase shifts for the lower parts are $0\pi$. The schematics in Fig.3(b) and Fig.3(c) illustrate the results of numerical modeling for the two selected positions of the external phase shifter $\Psi = 0.5\pi$ and $\Psi = 1.9\pi$. The power sensors depict the path for signal propagation, which is also shown by the dashed green line. There are no auto-oscillations if none of the paths meet the phase condition 4.2.

**Results of numerical modeling**

Multi-path active ring circuits can be applied for solving a variety of computational problems. The device structure may be modified for a specific task. Here, we present only three examples demonstrating possible practical applications and computing capabilities.

**Example 1: Prime factorization.** In Fig.4, there are shown the schematics of the multi-path circuit for prime factorization. It consists of a sequence of the two-path circuits similar to the one in Fig.2(a). The phase shifters for the upper paths in each block are adjusted to provide a phase shift proportional to the logarithms of prime numbers (e.g. $\pi \cdot \log(3)$; $\pi \cdot \log(5)$; $\pi \cdot \log(7)$; $\pi \cdot \log(11)$; and $\pi \cdot \log(13)$. All lower path phase shifters are set to provide zero phase shift. The bandpass filters are adjusted to allow only one frequency per the input-output path. For example, the signal on frequency $f_1$ can propagate only through the upper paths, the signal on frequency $f_2$ can propagate only through the first lower path and all other upper paths, the signal on frequency $f_n$ can propagate only through the lower paths. To factorize a given number $N$, the external phase shifter is set in the position $\Psi = 2\pi - \pi \cdot \log(N)$. The attenuator A is set to meet the amplitude condition 4.1 for all possible frequencies in the circuit. The phase condition for resonance (4.2) requires that the

$$\Psi + \sum_{i=1}^{5} \Delta_i = 2\pi, \tag{5}$$

where $\sum_{i=1}^{5}\Delta_i$ is the phase shift accumulated during signal propagation through the passive part. Only the signal coming through the path with a proper combination of the phase shifters will be amplified.

In Fig.5., there are shown the results of numerical simulations illustrating prime factorization. In Fig.5(a), it is shown the signal path for $N = 15$. There is only one path which matches the phase condition 4.2. The power sensors detect the self-oscillation power coming through the first and the second upper paths and the three lower paths. The phase shift accumulated by the signal is $\sum_{i=1}^{5}\Delta_i = \pi(\log(3) + \log(5))$. This is the only path that matches the phase condition 4.2. In Fig.5(b), there is shown the signal path for $N = 1001$. It goes through the first two lower paths and three upper paths. The phase shift accumulated by the signal is $\sum_{i=1}^{5}\Delta_i = \pi(\log(7) + \log(11) + \log(13))$. In Fig.5(c), there is shown the path for the maximum number $N = 15015$ that can be factorized with the given device. There is no self-oscillation path for $N = 107$ as there is no path providing the phase shift. We want to emphasize that the described circuit *can find any combination of multiple primes in one step*. In contrast, the Shor algorithm developed for quantum computers [5] can find only two primes at a time.

The multi-path part of the active ring circuit can be further extended to a 2D mesh. In Fig.6(A), it is schematically shown an active ring circuit where the passive part is a 3×3 matrix. The core of the matrix consists of nine delay lines connected to each other via the bandpass filters. The delay lines are marked as 1,2,…9. The signal can propagate only through horizontal and vertical waveguides as shown in Fig.6(A). There is a power sensor included in each delay line to detect the energy flow. Signal attenuation is directly related to the propagation distance. The distance between the nearest neighbor delay lines is $l_0$. The shortest input-output distance is $4l_0$. The longest path through the given matrix is $10l_0$. The collection of all possible paths connecting one input with one output can be found in the Supplementary Materials. The connection of the matrix with the external electric part is shown in Fig.6(B). Without a lack of generality, we have depicted just three inputs and three output ports on the left and the right sides of the matrix. Each input/output port has a switch that makes it possible to use all possible combinations of the input/output ports. There is a phase shifter and an attenuator at each output port. The ability to independently control signal phase and amplitude are important for a variety of search procedures.

**Example 2: Finding path(s) in a 2D grid with phase shifts matching the external phase.** In Fig.7 (a), it is shown a circuit with a given configuration of phase shifters and input/output connection. There are six delay lines with Δ= $0.1\pi$, and three delay lines with Δ= $0.2\pi$, Δ= $0.4\pi$, and Δ= $0.6\pi$. One input port #1 and all three output ports are connected to the matrix. The amplifier and the set of attenuators are set to provide the maximum amplification for all signal and satisfy condition 4.1 for all possible paths. We assume all three phase shifters to have the same value $\Psi_1 = \Psi_2 = \Psi_3 = \Psi$. The change in the external phase shifter makes the system search for a resonant path. It may be more than one path connecting input and output ports. In Fig.7(b), there are shown the results of numerical modeling on the number of paths for different positions of the external phase shifter. The external phase shifter is varied from $0\pi$ to $2\pi$ with a step of $0.1\pi$. There are no self-oscillations (i.e., resonant path) for external phase shifter $2\pi - \Psi < 0.6\pi$. Then, the first resonant path appears. It is shown in Fig. 7(C). the signal goes from input #1 to output #3 which is indicated by the green circles (i.e. power sensors). There may be a superposition of two, three, or even four paths for the other positions of the phase shifter $\Psi$. In Fig.7(D), there shown the results of numerical modeling for $\Psi = 2\pi - 1.1\pi$. There are two overlapping propagation paths connecting input #1 with outputs #1 and #3. There are two signals on two different frequencies propagating from the same input port to the different output ports. In Fig.7(E), there shown the results of numerical modeling for $\Psi = 2\pi - 1.8\pi$. There are four overlapping propagation paths (i.e. there are four signals on four frequencies) connecting input #1 with outputs #1 and #3.

**Example 3: Finding the shortest path for a given external phase.** The utilization of controllable attenuators allows us to extract the shortest part which exhibits a lower attenuation compared to other paths with the same phase shift. In Fig.8(A), it is shown the schematics of the circuit with a 3×3 matrix. The matrix is the same as in Fig.6. There is one input port # 1 and all three output ports connected to the matrix. The task is to find the shortest path for external phase $\Psi = 1.2\pi$. The external path is kept the same for all three outputs. There are four possible paths found in the previous example. To find the shortest one, the amplification is changed from the maximum to the minimum. Hereafter, we measure amplification in units $A_0$, such as $A_0 \cdot l_0 = 1$. The maximum of amplification of $10A_0$ makes all four possible paths to satisfy amplitude condition 4.1. Decreasing amplification to $7A_0$ leaves only three paths. Decreasing amplification to $4A_0$ leaves only one path. There are no self-oscillations for amplification levels below $4A_0$ as the gain coming from the amplifier is no sufficient to compensate signal damping during the propagation. In Fig.8(B), the number of paths is shown for different amplification levels. It starts from four and decreases to zero for different levels of amplification. The green circles in Fig.8 (C-E) illustrate the outputs of the power sensors which reveal the signal path.

**Example 4: Finding the shortest path through specific points in the mesh.** In Fig.9(A), there is shown a matrix where six delay lines are chosen to provide phase shift $0\pi$, while delay lines #2,#4, and #6 provide phase shift $0.5\pi$, $0.3\pi$, and $0.7\pi$, respectively. The task is to find the shortest path connecting input #1 to output #3 and coming through delay lines #2,#4, and #6. To find the path, the external phase shifter at output port #3 is set to $0.5\pi$. Next, the amplification is varied from the maximum of $10A_0$ to the minimum $4A_0$. All power sensors detect energy flow for maximum amplification of $10A_0$ as illustrated in Fig. 9(B). There are only seven sensors detecting energy flow left for $8A_0$. There are no self-oscillations in the system for lower amplification levels. The key idea of the search procedure is to set the output phase to the sum of phase delays of given delay lines to ensure the signal goes through the points of interest, and change the attenuation from maximum to minimum. The set of power sensors will depict the shortest path through the selected sites. This example shows the capabilities of the combinatorial devices for solving the Seven Bridges of Königsberg problem. It is a historically notable mathematical problem. Its negative resolution by Leonhard Euler in 1736 laid the foundations of graph theory and prefigured the idea of topology [13]. The city of Königsberg in Prussia (now Kaliningrad, Russia) was set on both sides of the Pregel River, and included two large islands - Kneiphof and Lomse - which were connected to each other, or to the two mainland portions of the city, by seven bridges. The problem is to devise a walk through the city that would cross each of those bridges once and only once. The bridges can be represented by the delay lines providing unique phase shifts (e.g., logarithms of prime numbers). Other delay lines in the mesh are set up to provide zero phase shift. The system will naturally search for the path that matches the sum of phase shifts of the given delay lines (cities). The shortest path is then extracted out of many by reducing the amplification level. The same approach can be applied to solving the traveling salesman problem.

**Experimental demonstration**

All the presented above examples are based on the ability of active ring circuits to self-adjust to the resonant (i.e., self-oscillation) frequency. Our approach is to utilize this physical phenomenon for finding the path out of many possible which matches the self-oscillation conditions 4.2 To validate this idea, we accomplished a proof-of-the-concept experiment for a two-path active ring circuit. The schematics of the device are shown in Fig.10. it is an active ring circuit with two delay lines made of YIG. YIG is chosen due to the low spin wave damping, which makes it possible spin wave propagation on larger distances (e.g., up to 1cm) at room temperature. Delay line one is based on YIG-film with thickness $d_0$ 9.6 = μm, width 2 mm, length 9 mm, saturation magnetization $4\pi M_0$ = 1750 G. Delay line two is based on YIG-film with thickness $d_0$ = 21.3 μm, width 6 mm, length 12 mm, saturation magnetization $4\pi M_0$ = 1750 G. The two waveguides are attached to a permanent magnet based on K&J Magnets, model BX8X84. There is a set of three direction couplers DC-0, DC-1, and DC-2. A small portion of the oscillation power is taken by the coupler to check the energy flow in the different parts of the circuit. For example, DC-0 coupler allows us to check the total power in the circuit. DC-1 and DC-2 are aimed to detect energy flow through the upper and lower paths, respectively.

Spin wave dispersion depends on the strength and the direction of the bias magnetic field [14]. For instance, spin waves propagating perpendicular to the external magnetic field, *i.e.* magneto-static surface spin waves (MSSWs) and the spin waves propagating parallel to the direction of the external field, *i.e.* backward volume magneto-static spin waves (BVMSWs) possess significantly different dispersion. The relation between the frequency and the vector are the following [15]:

$$f_{MSSW} = \sqrt{\left(f_H + \frac{f_M}{2}\right)^2 - \left(\frac{f_M}{2}\right)^2 exp(-2kd_0)}, \qquad (6)$$

$$f_{BVMSW} = \sqrt{f_H \left(f_H + f_M \frac{1-\exp(-kd_0)}{kd_0}\right)},$$

where $f_H = \gamma H_0$, $f_M = 4\pi\gamma M_0$, $k$ is the wavenumber, $d_0$ is the film thickness. The difference in the dispersion manifests itself not only in the different phase shifts accumulated for the same propagation length but also in the frequency interval for propagating spin waves. Thus, magnetic waveguides can be utilized as delay lines and frequency filters at the same time. It makes spin wave delay lines convenient for application in the proposed multi-path devices. As we mentioned before, relatively low spin wave velocity makes it possible to obtain prominent phase shifts to RF signals on sub-millimeter propagation distances. The frequency window for propagating spin waves and the phase shift accumulated in a given waveguide can be controlled by the external magnetic field. The first spin wave logic device built by Kostylev et al., in 2005 [15] utilized an electric current-carrying wire under the YIG film to control the phase shift accumulated by the propagating spin wave. In our experiment, we used BVMSW configuration for the spin wave propagating the upper path and MSSW configuration for spin waves propagating in the lower path.

Prior to the proof-of-the-concept experiment, spin wave transport was studied in each delay line separately. Experimental data are shown in Fig.11. The black curve corresponds to the spin wave signals propagating in Delay line -1 in BVMSW configuration. The red curve corresponds to the spin wave signals propagating in Delay line -2 in MSSW configuration. The data are measured by Programmable Network Analyzer (Keysight N5241A). The graph shows $S_{21}$ parameter in dB in the frequency range from 2.2 GHz to 2.9 GHz. There are two peaks at frequencies around 1.983 GHz and 2.208 GHz for delay lines 1 and 2, respectively. The main portion of the power of spin wave signal through DL-1 is coming on 1.983 GHz. The main portion of power of spin wave signal through DL-2 is coming on 2.208 GHz. It should be noted that these prelim data were obtained for magnetic waveguides not connected in the active ring circuit and not amplified.

Next, the two delay lines were connected in the active ring circuit as shown in Fig.10. In Fig.12(A), there are shown experimental data on the power circulating in the ring circuit measured through DC-0. The measurements are taken at the different positions of the external phase shifter $\Psi$. The data show prominent power variation depending on the external phase. The phase is shown in div units which are frequency dependent. One div at 1.983 GHz corresponds to 10 degrees, while one div at 2.208 GHz corresponds to 12 degrees. There are no auto-oscillations in the range 9 div to 12 div. ($120^0$ to $160^0$). All measurements are accomplished at the fixed amplification of 18 dB, which is just below the self-oscillation threshold. In Fig.12(B), there are shown experimental data on the dominant frequency (i.e., the frequency caring the most portion of energy) of circulating self-oscillations depending on the position of the external phase shifter. The data are obtained using Spectrum Analyzer GW Instek, model GSP 827 for signal taken at DC-0. As expected, there are two distinct frequencies 1.983 GHz and 2.208 GHz which is consistent with the experimental data shown in Fig.10.

Finally, the energy flow through the upper and lower paths was measured separately using DC-1 and DC-2 with the help of microwave detector KEYSIGHT 33330B connected to KEITHLEY 2182A nanovoltmeter. In Fig.13(A), there are shown experimental data on the power $A_s$ [mV] depending on the position of the external phase shifter. The black markers correspond to the signal propagating through delay line 1. The red markers correspond to the signal propagating through delay line 2. In Fig.13(B), there is shown the ratio in dB between the power flow flowing through the upper and lower paths. The difference in the power flow between the paths exceeds 20 dB for all self-oscillation regions. The maximum difference exceeds 40 dB for the external phase from 12 to 22 divisions ($162^0$ to $300^0$) where most of the power comes through the delay line 1. It takes less than 1 μs for the system to come to the stable self-oscillation. All experiments are done at room temperature.

**Discussion**

There are several observations we would like to outline based on the obtained experimental data. (i) The switching of the power flow between the paths in the magnetic matrix is observed depending on the external phase shifter. This fact itself provides an intriguing possibility of controlling energy flow in micro- and nano-structures via an external parameter – phase. The utilization of an external (relatively large) phase shifter for signal re-direction (e.g., on the sub-micrometer scale) may be useful for nanostructure devices addressing/control. (ii) The described combinatorial logic devices are digital. The output (i.e., the results of computation) is detected by the set of power sensors. The difference between energy-carrying and non-carrying paths may exceed 40 dB at room temperature. It is on the same scale as the On/Off ratio for modern transistors. Active ring circuits can be integrated with conventional electronic devices and complement CMOS in the special task data processing. (iii) There may be a variety of practical implementations of the active ring circuits with a multi-path passive part using different materials and signal delay mechanisms. Spin wave delay lines are convenient due to the small size and ability to control dispersion by the bias field. For instance, a voltage-controlled spin wave modulator based on the synthetic multiferroic structure was recently demonstrated [16].

To estimate the functional throughput of the proposed device, let us consider a circuit comprising an $n \times n$ matrix with $n$ inputs and $n$ output ports. There are $n^2$ delay lines, $n^2$ frequency filters, and $n^2$ power sensors. The schematics of a 5×5 matrix are shown in Fig.14. The signal can propagate on horizontal, vertical, and diagonal waveguides connecting the nearest-neighbor sites. Each site includes a delay line and a frequency filter. There is a set of *s* distinct operational frequencies (i.e. the frequencies at which the bandpass filters can transmit signals). There are $2^s$ possible phase combinations per filter. There may be filters with the same set of passing frequencies. In general, there are $(n \times n)!$ possible filter arrangements in the matrix. It should be noted that the total number of frequencies is much less than the number of paths in the mesh and less than the number of filters: $s < n \times n \ll n!$ The position of filters as well as the set of passing frequencies is fixed and does not change during the device operation. There are $n$ phase shifters and $n$ attenuators at the output. There is just one amplifier in the device.

The principle of operation is the following. The phase shifts per delay lines are adjusted by the external parameter (e.g., electric field) prior to computation. There is no restriction on the phase shift values. For instance, it may be a set of unique phase shifts as in Examples 1 and 4, or a combination of unique and identical shifts as in Examples 2 and 3. There are $2n$ switches. There are two positions for a switch: On and Off. There is a number of connection combinations. For example, one input –one output combination, two input-one output port, etc. There are $2^{2n-2}$ possible combinations for the $2n$ input/output switches. There should be at least one input and one output port connected to the electric part. A set of phase shifters together with input/output switch combination can be considered as a database to be searched. The number of possible paths between the given input/output ports and set of phase shits depends on the size of the matrix. For example, the number of paths connecting the most distant ports (i.e., input port #1 and the output port # $n$) can be calculated as follows[17]:

$$Number\ of\ paths\ (1\ input - 1\ output) =\ (n+n)!/(n! \times n!) \tag{7}$$

For example, there are 252 possible paths connecting the most distant ports for the 5×5 matrix shown in Fig.14. The number of possible paths just between the most distant ports in 50×50 matrix exceeds $10^{29}$. The total number of paths for all possible combinations of input and output ports can be estimated as follows:

$$Total\ number\ of\ paths\ \approx 2^{2n-2} \cdot (n+n)!/(n! \times n!) \tag{8}$$

The first term on the right corresponds to the number of input-output port combinations, while the second term is related to the number of paths between the ports. In general, the number of paths varies for different ports. For simplicity, the number of paths is taken the same to all port combinations according to Eq.(7).

The instructions are encoded into the position of the output external phase shifters and attenuators. There are $z^n$ possible phase shifter combinations, where $z$ is the number of possible phase positions per shifter. The number of distinct and recognizable phases $z$ is limited by the accuracy of the phase shifter. For instance, the accuracy of the phase shifters used in the experimental part ARRA 9428A is about 2 degrees. It can be taken as a practical benchmark, which limits the number of possible phases per shifter to 180. The number of possible attenuation levels for the attenuator depends on the size of the mesh. For instance, the maximum amplification is related to the signal attenuation through the longest possible path. The minimum amplification is required for the shortest path connected the input and output ports. For example, the minimum and maximum values are $nA_0$ and $n^2A_0$ for a $n \times n$ 2D mesh. It would be reasonable to consider 20 distinct amplification levels (3 dB difference, 60 dB range) for the benchmark. The total number of instructions can be estimated as follows:

$$Number\ of\ instructions = 2^{n-1} \cdot 180^n \cdot 20^n, \tag{9}$$

where the first term on the right corresponds to the number of combinations for output switches, the second term on the right is the number of possible phase combinations among $n$ output phase shifters, and the last term on the right is the number of amplitude level combinations.

As soon as the matrix is connected to the external electric circuit, it starts the search for a resonant path. The result of the computation (i.e., the path) is detected by the set of power sensors. The power level is compared to some reference level $W_0$. The logic output is 1 if $W_k \geq W_0$, and the output is 0 otherwise. Thus, the output is a sequence of $n^2$ bits (e.g., 100010..001) which describes the path of signal propagation. The area of the combinatorial device is mainly defined by the size of the multi-path matrix. It can be estimated as $l^2 \cdot n^2$, where $l$ is the characteristic size of the mesh cell comprising a delay line and four bandpass filters. The computational time depends on the size of the mesh and the signal group velocity $l \cdot n^2 / v_g$. For spin waves, the characteristic size can be scaled down to tens of nanometers, while the group velocity is about $10^4$ m/s. The overall, functional throughput of combinatorial logic devices can be estimated as follows:

$$Functional\ throughput\ (combinatorial\ logic) = \frac{2^{2n-2} \cdot (n+n)!/(n! \times n!)}{l^2 \cdot n^2 \times l \cdot n^2 / v_g}. \tag{10}$$

Regardless the size of the delay line and signal propagation speed, the functional throughput of the combinatorial logic devices increases proportionally to $n$ factorial. For example, taking $l = 100\ \mu m$ and $n = 50$, one can estimate the area of the device to be $25\ mm^2$, the computational time is $25\ \mu s$. The functional throughput exceeds $10^{60}$ operations per meter squared per second. It will take a bit longer than the age of the universe (13.77 billion years) to check one-by-one all paths (one path per ns) for current supercomputers.

There is some similarity between the operation of Quantum computers and combinatorial logic devices. In both cases, the computation starts with a superposition of states. It is a superposition of quantum qubits in a quantum computer. It is a superposition of signals propagating in all possible paths in a combinatorial device. In quantum computer, the searched combination appears with a higher probability in the output. In combinatorial logic, the searched path (i.e. energy flow through the searched path) is amplified. Potentially, combinatorial logic devices may compete with quantum computers in functional throughput by searching through a larger number of possible combinations for the same number of inputs $n$.

Combinatorial devices can be also utilized for information storage. The number of paths increases factorial with the size of the mesh. It constitutes a fundamental advantage over conventional memory. The advantage lies in the fact that the output of the combinatorial device depends on the mutual arrangement of the delay lines (i.e., memory elements) in the mesh as well as the combination of the input-output ports. To comprehend the advantage in the storage capacity, one can compare the number of queries and bits retrieved from the conventional magnetic 5×5 memory and 5×5 combinatoric device. One query for conventional memory is associated with magnetoresistance measurement for a selected magnetic bit. There are two lines named "word" and "bit" connected simultaneously to address the memory element on the intersection. Other input/output combinations are not informative and not utilized. The result may be either high or low resistance compared to the reference one (i.e., one bit). Thus, there are 25 possible quires with one bit per query retrieved for conventional memory. The number of queries for the combinatorial device is defined by the number of instructions given by Eq.(9). Each query is a path measurement, where one path is described by 25 bits (i.e., the On/Off of the 25 power sensors). There are 921600 queries with 25 bits per query for the combinatorial device.

There are certain physical restrictions on the number of operational frequencies, phase shifts, and attenuation levels that can be practically implemented. The most critical concern is related to the number of operational frequencies (i.e., the number of passing frequencies per bandpass filter). For example, each path in Example 1 is associated with a distinct frequency. The number of possible paths increases exponentially with the number of two-path blocks. However, the number of operational frequencies is limited in any practical device. Also, the bandpass filters in Examples 2 and 3 should be arranged in such a way that there is only one frequency for one input-output path. At the same time, *the same frequencies can be utilized for not-crossing paths*. For example, the same frequency $f_1$ can be used for paths 1-4-7, 2-5-8, and 3-6-9 (see Fig.6(A)). It leads to a combinatorial problem of arranging a number of bandpass filters with a limited number of operational frequencies in the mesh to ensure the maximum number of not-crossing paths. The number of phase and amplitude levels is limited by the precision of the hardware. More critical is the increase of the number of possible amplitude levels with the size of the matrix. This fact will limit the range of paths that can be checked. There are other important questions related to the delay line dispersion, the width of resonance (e.g., spin wave resonance), design rules in the multi-path matrix, etc. This work is aimed to present the concept of combinatorial logic devices and outline their most appealing properties.

To conclude, we described the concept of combinatorial logic devices where an act of computation is associated with finding a path(s) for signal propagation. It is based on an active ring circuit with a multi-path magnetic matrix. The system naturally searches for a resonant path that matches the self-oscillation conditions. Combinatorial logic devices may compliment CMOS-based computers in special task data processing. There are several examples illustrating prime factorization and finding the shortest path connecting giving parts on the 2D mesh. There are also presented experimental data showing signal path change depending on the external phase shifter. The proof-of-the-concept experiment is accomplished with two spin wave delay lines. The direction of the bias magnetic field is chosen to provide different frequency and phase shifts to propagating spin wave signal. It is shown the switching of propagation paths depending on the position of the external phase shifter. The difference in the power flow exceeds 40 dB at room temperature. The combinatoric device can be utilized for information storage by providing a larger capacity compared to conventional memory. These devices can potentially compete with quantum computers in functional throughput.

## Methods

**Device Fabrication**

The delay lines are made of single-crystal $Y_3Fe_2(FeO_4)_3$ films. The films were grown on top of a (111) Gadolinium Gallium Garnett ($Gd_3Ga_5O_{12}$) substrate using the liquid-phase epitaxy technique. The micro-patterning was performed by laser ablation using a pulsed infrared laser ($\lambda \approx 1.03$ μm), with a pulse duration of ~256 ns. Delay line one is based on YIG-film with thickness $d_0$ 9.6 = μm, width 2 mm, length 9 mm, saturation magnetization $4\pi M_0$ = 1750 G. Delay line two is based on YIG-film with thickness $d_0$ = 21.3 μm, width 6 mm, length 12 mm, saturation magnetization $4\pi M_0$ = 1750 G.

**Measurements**

There are two П-shaped micro-antennas fabricated on the edges of the delay line. Antennas were fabricated from a gold wire of thickness 24.5μm and placed directly at the top of the YIG surface. One of the antennas is used to generate spin wave signal. The other antenna is to detect the inductive voltage produced by the spin wave signal. The set of attenuators (PE7087) and phase shifters (ARRA 9428A) is used to control the amplitudes and the phases in each delay line.

**Author Contributions**

A.K. conceived the idea of combinatorial logic gates, provided numerical modeling, and wrote the manuscript, M.B. carried out the experiments. All authors discussed the data and the results, and contributed to the manuscript preparation.

**Competing financial interests**

The authors declare no competing financial interests.

**Data Availability Statement**

The data that support the findings of this study are available from the corresponding author upon reasonable request.

**Supplementary Material**

The supplementary material section contains data to support Examples 2 and Example 3.

**Acknowledgment**

This work was supported in part by the INTEL CORPORATION (Award #008635, Spin Wave Computing) (Project director is Dr. D. E. Nikonov).

**Figure Captions**

Figure 1. (A) Schematics of an active ring system comprising a broadband amplifier G, a magnonic waveguide – a delay line providing phase shift Δ, a variable attenuator A, and a variable phase shifter Ψ connected in series. The connection between the parts is via coaxial cables. Two microstrip antennas excite and receive spin waves through the waveguide. The signal circulating in the ring circuit exibits a conversion from electromagnetic wave to spin waves and vice versa. Auto-oscillations occur when the electric and magnetic parts match each other. (B) Results of numerical modeling illustrating the amplitude evolution with a number of rounds for different phase correlations. The black curve corresponds to the resonant conditions where the total phase shift accumulated through a round of propagation $\Psi + \Delta = 2\pi$. The circulating signals are coming in-phase and enhance the amplitude. The red and the blue curves corresponds to the cases $\Psi + \Delta = 0.1\pi$ and cases $\Psi + \Delta = 0.3\pi$, respectively.

Figure 2. (A) Schematics of an active ring circuit with two passive paths. Each path includes a bandpass filter, a delay line providing a phase shift Δ to the propagating signal, and a power meter. The power meter is aimed to detect whether the power of the signal flowing through the path exceeds the noise level. (B) Results of numerical modeling showing the amplitude of the auto-oscillations depending on the position of the external phase shifter $\Psi$. The attenuator A is set to provide maximum amplification to meet condition 4.1. The phase shits provided for the upper and the lower paths are $\Delta_1 = 1.7\pi$ and $\Delta_2 = 1.0\pi$, respectively. The auto-oscillations take place for the two positions of the external filter at which condition 4.2. is satisfied. The most of the power goes through the upper path for external phase $\Psi = 0.3\pi$. Most of the power goes through the lower path for $\Psi = 1.0\pi$. The green and the red circles depict the output of the power sensors (i.e., green circle means power, red color means no power).

Figure 3. (a). Schematics of the device comprising three two-path blocks. The phase shifts Δ for the upper paths are $0.1\pi$, $0.3\pi$, and $1.5\pi$, respectively. The phase shifts for the lower parts are $0\pi$. The power sensors are depicted as circles embedded in the delay lines. The set of pass frequencies is shown for each bandpass filter. (B) Results of numerical modeling illustrating signal propagation for $\Psi = 0.5\pi$. The propagation path is depicted by the green dashed line. (C) Results of numerical modeling illustrating signal propagation for $\Psi = 0.2\pi$. The green circle corresponds to the energy flow through the delay line.

Figure 4. Schematics of the device for prime factorization. The delay lines in the upper paths provide phase delays proportional to the logarithms of prime numbers 3,5,7,11, and 13, respectively. The delay lines in the lower paths provide no phase shift for the propagating signal. The external phase shifter $\Psi$ is setup to $\Psi = 2\pi - \pi \cdot \log(N)$, where $N$ is the number to be factorized.

Figure 5. Results of numerical modeling on prime factorization. The green circles depict energy flow through the delay line. Red circles depict no self-oscillation power coming through the delay line. (A) Sensors output for $N = 15$. (B) Sensors output for $N = 1001$. (C) Sensors output for $N = 15015$. This is the largest prime number that can be factorized with the given circuit. (D) Sensor output for $N = 107$. All power sensors are shown red. There are no self-oscillations in this case.

Figure 6. (A). Schematics of 3×3 magnetic matrix comprising nine delay lines marked as 1,2,…9. The delay lines are connected through the bandpass filters. The signal can propagate only through horizontal and vertical waveguides. There is a power sensor included in each delay line to detect the energy flow. (B) Matrix connection to the electric part. There are three input and three output ports. Each port has a switch. There are phase shifters and variable attenuators connected to each output port.

Figure 7(A). Schematics of the 3×3 matrix with a given set of phase shifts per delay line. Delay lines #1,#2,#3,#7,#8,and #9 provide phase shift with $\Delta = 0.1\pi$. Delay lines #4,#5 and #6 provide phase shifts $\Delta = 0.2\pi$, $\Delta = 0.4\pi$, and $\Delta = 0.6\pi$, respectively. One input port #1 and all three output ports are connected to the matrix. The amplifier and the set of attenuators are set to provide the maximum amplification for all signals and satisfy condition 4.1 for all possible paths. All three phase shifters have the same value $\Psi_1 = \Psi_2 = \Psi_3 = \Psi$. (B) Results of numerical modeling on the number of paths for different positions of the external phase shifter. The external phase shifter is varied from $0\pi$ to $2\pi$ with a step of $0.1\pi$. There are no self-oscillations (i.e., resonant path) for external phase shifter $2\pi - \Psi < 0.6\pi$. (C) Results of numerical modeling illustrating signal propagation path for $\Psi = 2\pi - 0.6\pi$. Green circles for power sensors depict power. The path is also shown by the dashed green line. There is only one path for the selected position of the external phase shifter. (D) Results of numerical modeling for $\Psi = 2\pi - 1.1\pi$. There are two overlapping propagation paths connecting input #1 with outputs #1 and #3. (E) Results of numerical modeling for $\Psi = 2\pi - 1.8\pi$. There are four overlapping propagation paths connecting input #1 with outputs #1 and #3.

Figure 8. (A) Schematics of the 3×3 matrix with a given set of phase shifts per delay line. The set is the same as in Fig.7(A). There is one input port # 1 and all three output ports connected to the matrix. The external phase is fixed to $\Psi = 1.2\pi$ for all three outputs. The task is to find the shortest input-output path for the given phase. (B) Results of numerical modeling showing the number of paths as a function of amplification level. There may be four, three, one, and no paths. (C-E) Results of numerical modeling illustrating power sensor output (i.e. green circles) for the different levels of amplification. The shortest path appears for amplification $4A_0$ as shown (E).

Figure 9. (A) Schematics of the matrix where six delay lines are chosen to provide phase shift $0\pi$, while delay lines #2,#4, and #6 provide phase shift $0.5\pi$, $0.3\pi$, and $0.7\pi$, respectively. The task is to find the shortest path connecting input #1 to output #3 and coming through delay lines #2,#4, and #6. (B). Results of numerical modeling showing the possible paths for the maximum amplification of $10A_0$. All power sensors detect energy flow. (C) Results of numerical modeling showing the path for amplification of $8A_0$. There are only seven sensors detecting energy flow. That is the only and the shortest path through the selected sites.

Figure 10. Schematics of the experimental setup for the proof-of-the-concept experiment with a two-path circuit. It is an active ring circuit with two delay lines made of YIG. Delay line one is based on YIG-film with thickness $d_0$ = 9.6 µm, width 2 mm, length 9 mm, saturation magnetization $4\pi M_0$ = 1750 G. Delay line two is based on YIG-film with thickness $d_0$ = 21.3 µm, width 6 mm, length 12 mm, saturation magnetization $4\pi M_0$ = 1750 G. The two waveguides are attached to permanent magnet based on K&J Magnets, model BX8X84. There are three direction couplers DC-0, DC-1, and DC-2. DC-0 coupler is to measure the total power in the circuit. DC-1 and DC-2 are aimed to detect energy flow through the upper and lower paths, respectively.

Figure 11. Experimental data on spin wave transport through the delay lines. The measurements are accomplished for each delay line separately. The delay lines are not connected to the active ring circuit. The graph shows $S_{21}$ parameter in dB in the frequency range from 2.2 GHz to 2.9 GHz. The black curve corresponds to the spin wave signals propagating in Delay line -1 in BVMSW configuration. The red curve corresponds to the spin wave signals propagating in Delay line -2 in MSSW configuration.

Figure 12. Experimental data measured at DC-0. (A) Circulating power at different positions of the external phase shifter $\Psi$. There are no auto-oscillations in the range 9 div to 12 div. ($90^0$ to $120^0$). All measurements are accomplished at the fixed amplification of 18 dB, which is just below the self-oscillation threshold. (B) The dominant frequency of the energy flow. The data are obtained using Spectrum Analyzer. There are two dominant frequencies consistent with the results of preliminary testing.

Figure 13. Experimental data on the energy through the upper and lower paths measured separately using DC-1 and DC-2. The black markers correspond to the signal propagating through delay line 1. The red markers correspond to the signal propagating through delay line 2. (A) Experimental data on the power $A_s$ [mV] depending on the position of the external phase shifter. (B) The ratio in dB between the power flow flowing through the upper and lower paths. All experiments are done at room temperature.

Figure 14. Schematics of a 5×5 matrix. The signal can propagate on horizontal, vertical, and diagonal waveguides connected to the nearest-neighbor sites. Each cite includes a delay line, frequency filter, and power detector. There are five inputs and five outputs on the two sides of the matrix. Each port has a switch to control connectivity. There is one amplifier. There are 562 paths connecting input port 1 and output port 5. The green circles depict the superposition of three selected paths. There are $2^{25}$ possible outputs of the power sensor. All of them can be recognized.

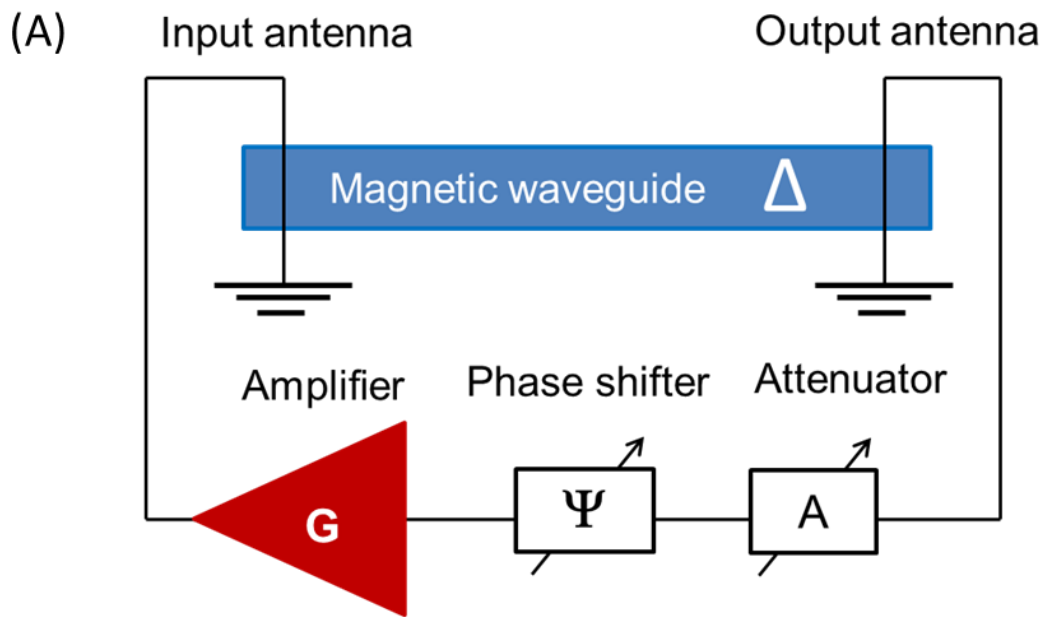
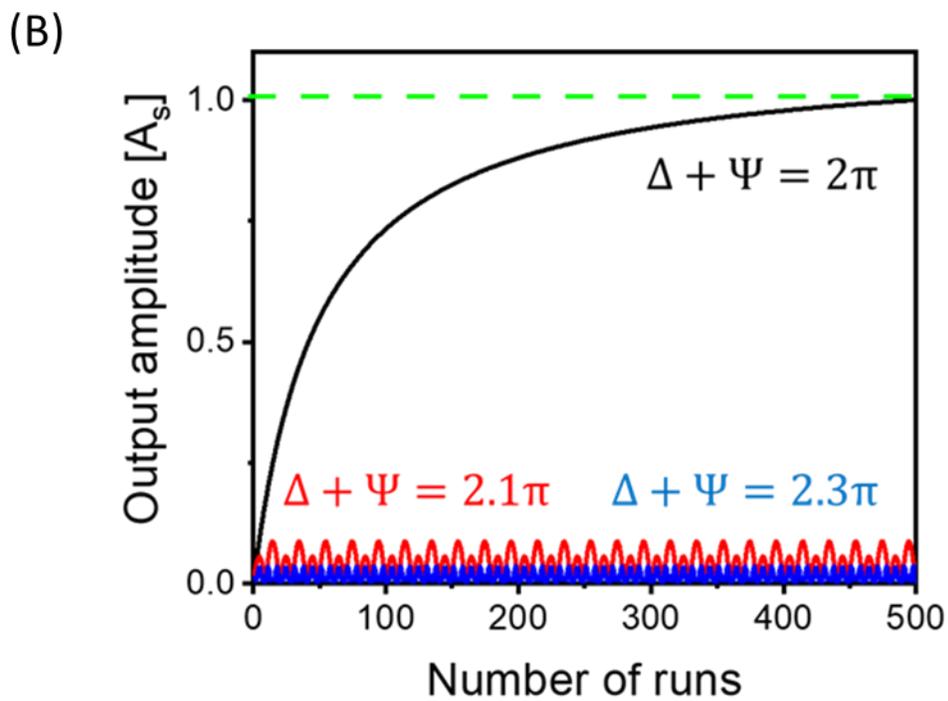

**Figure 1**

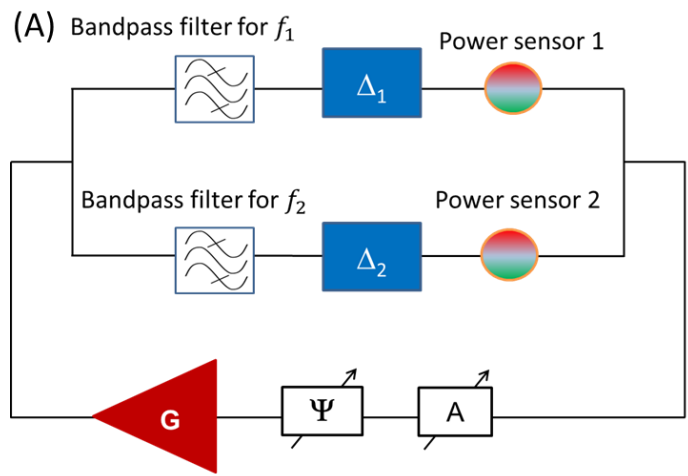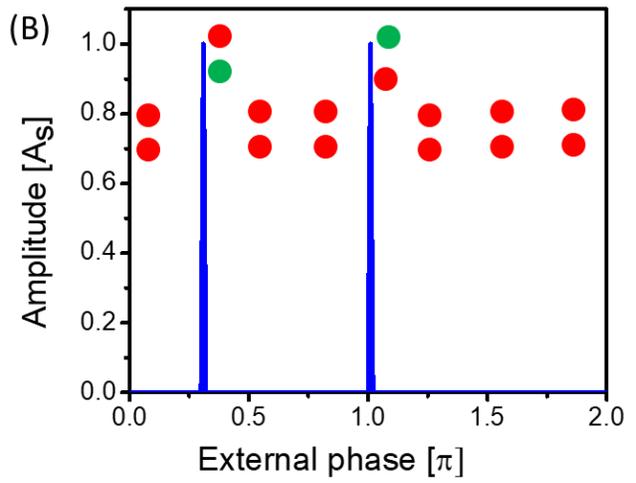

**Figure 2**

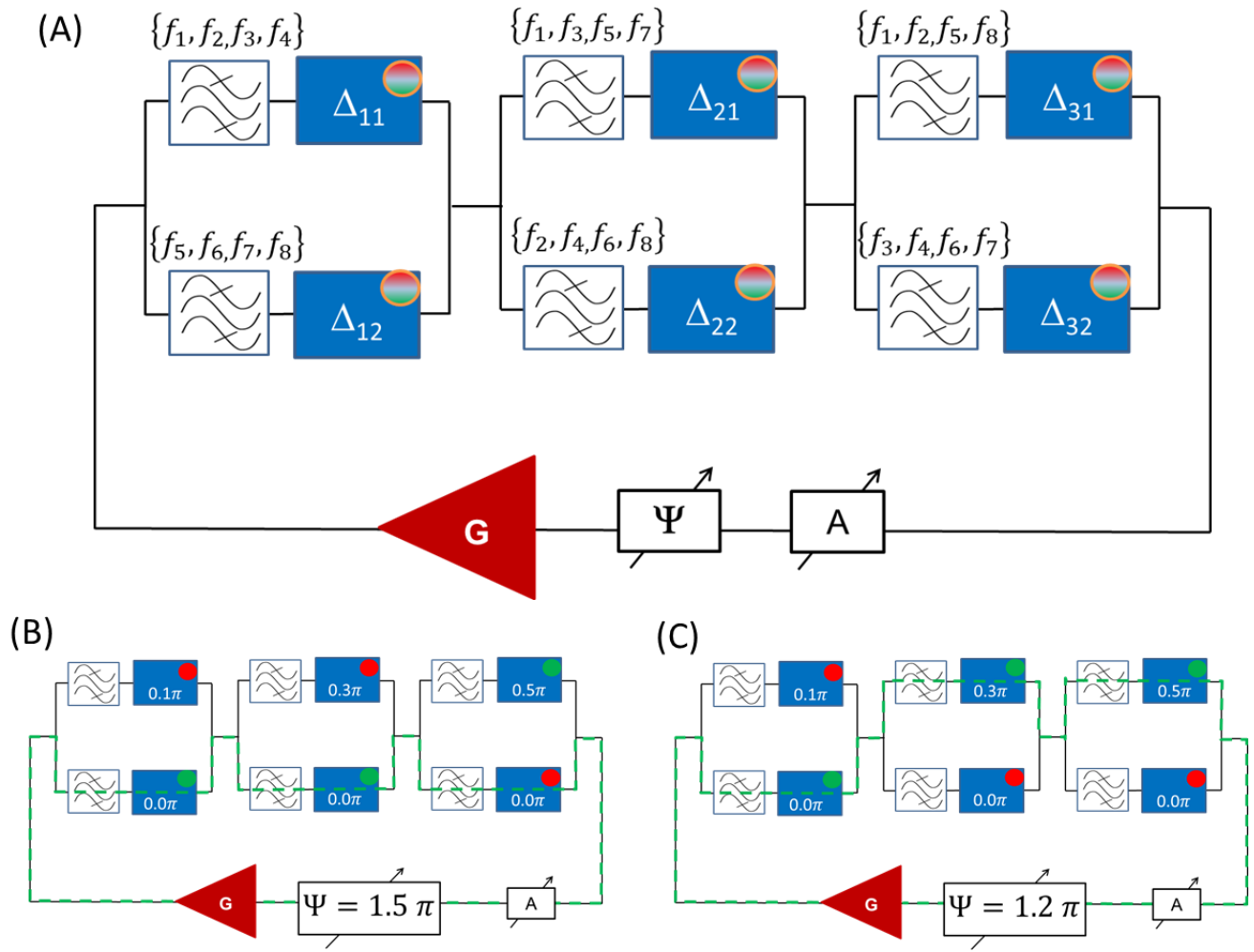

**Figure 3**

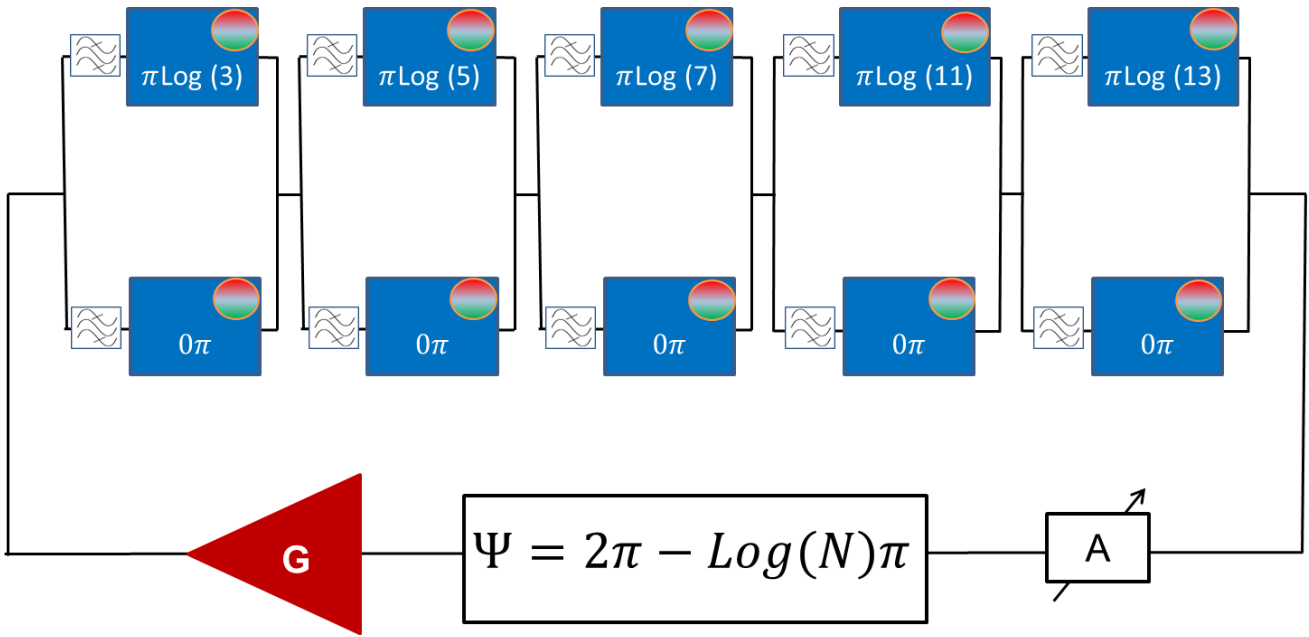

**Figure 4**

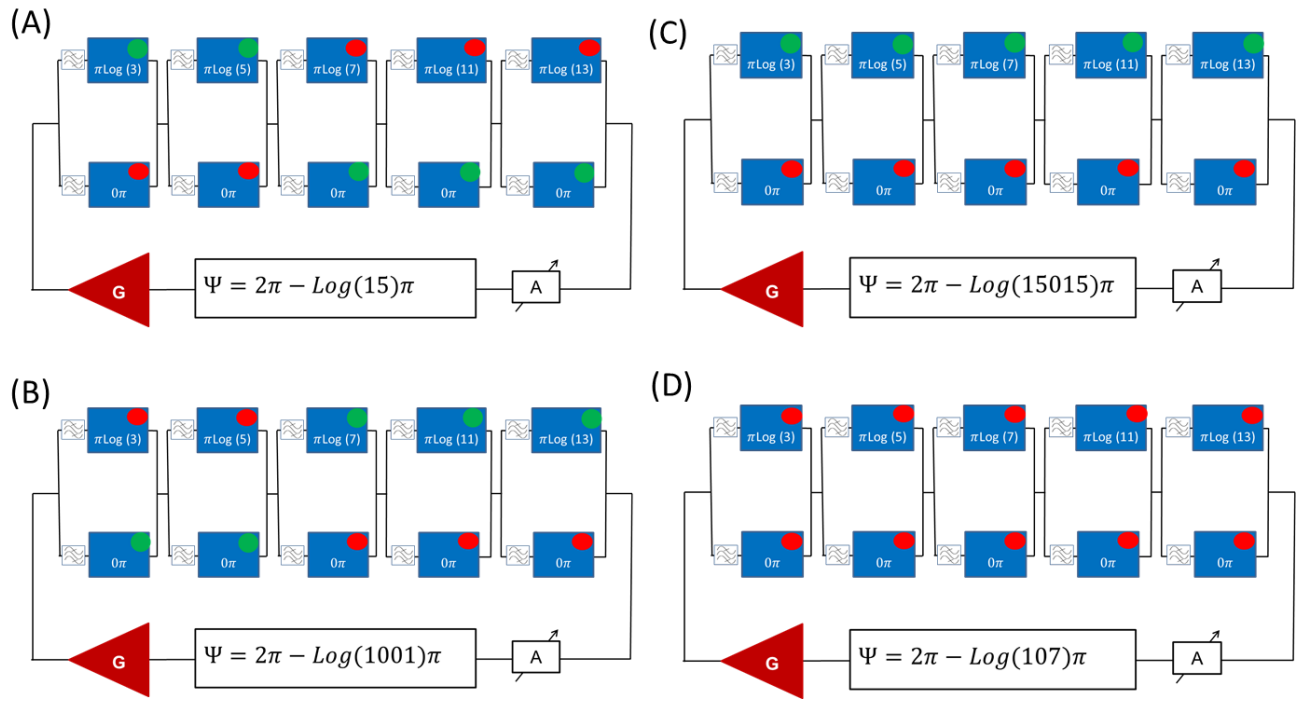

**Figure 5**

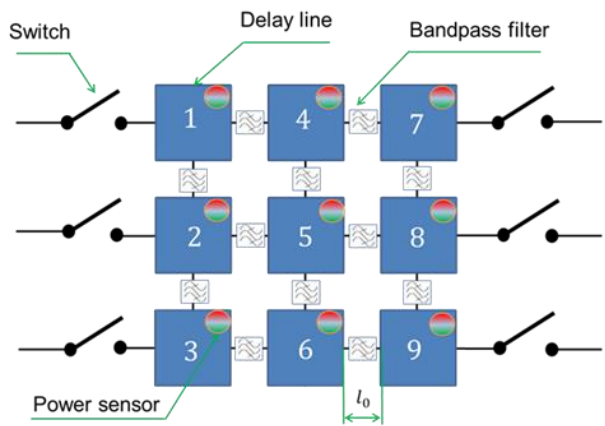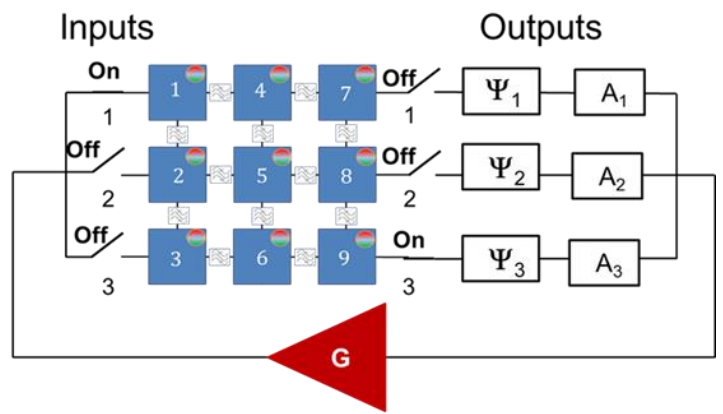

**Figure 6**

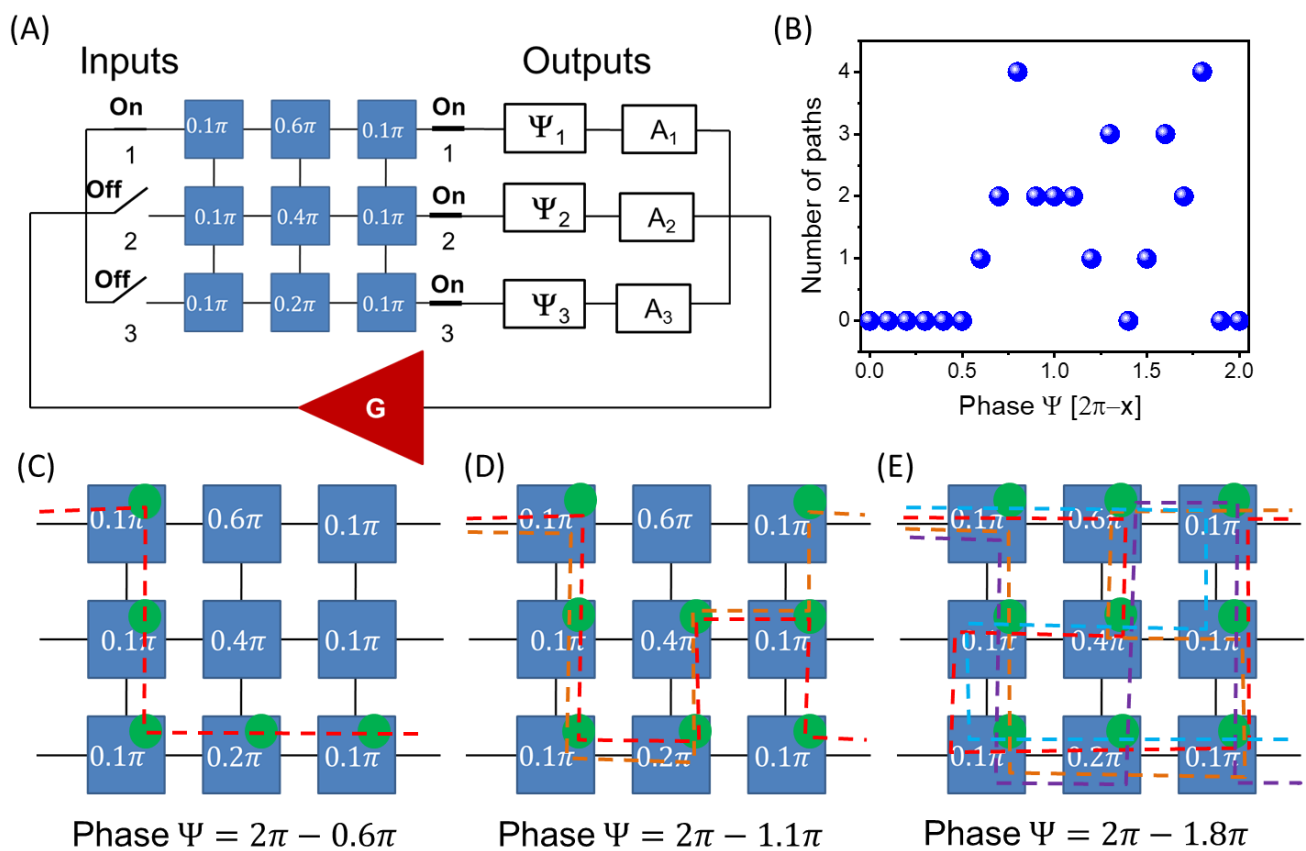

**Figure 7**

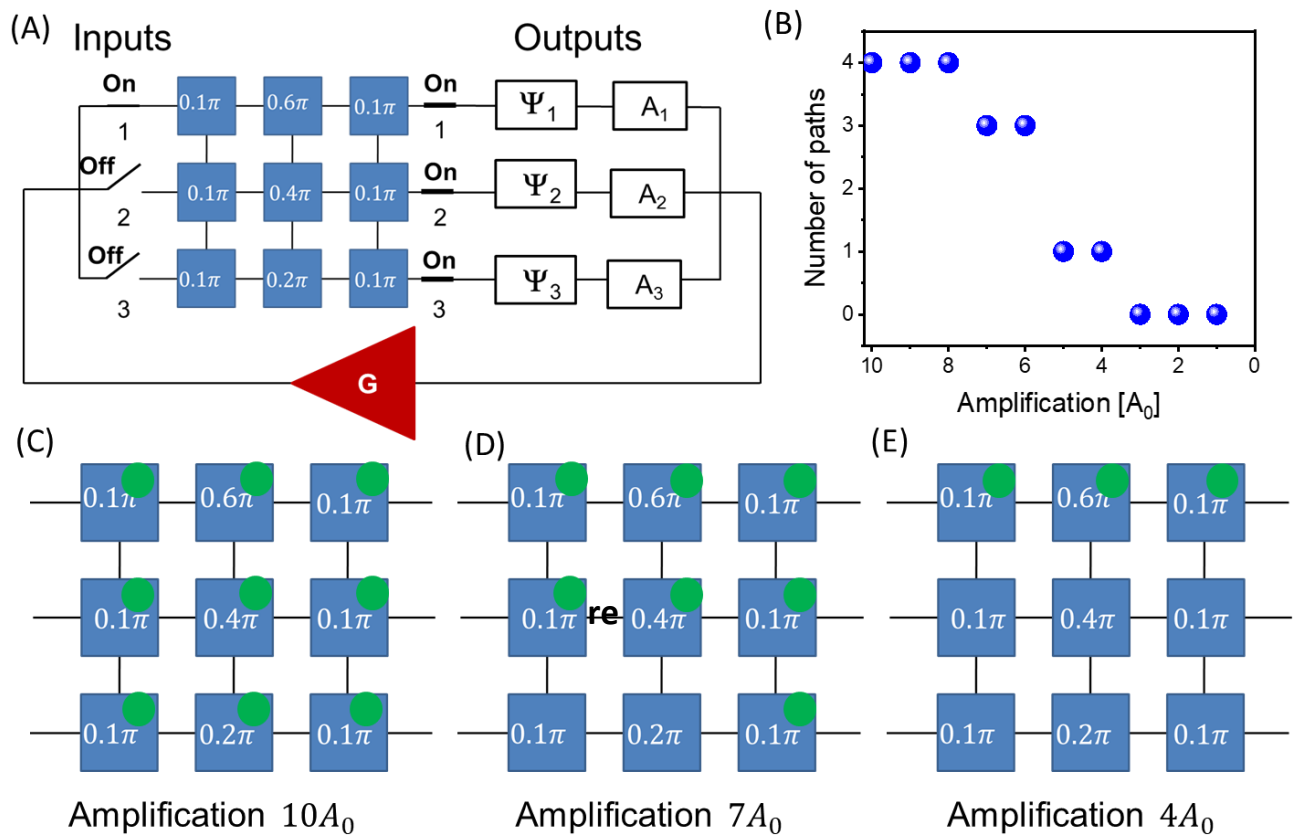

**Figure 8**

**Figure 9**

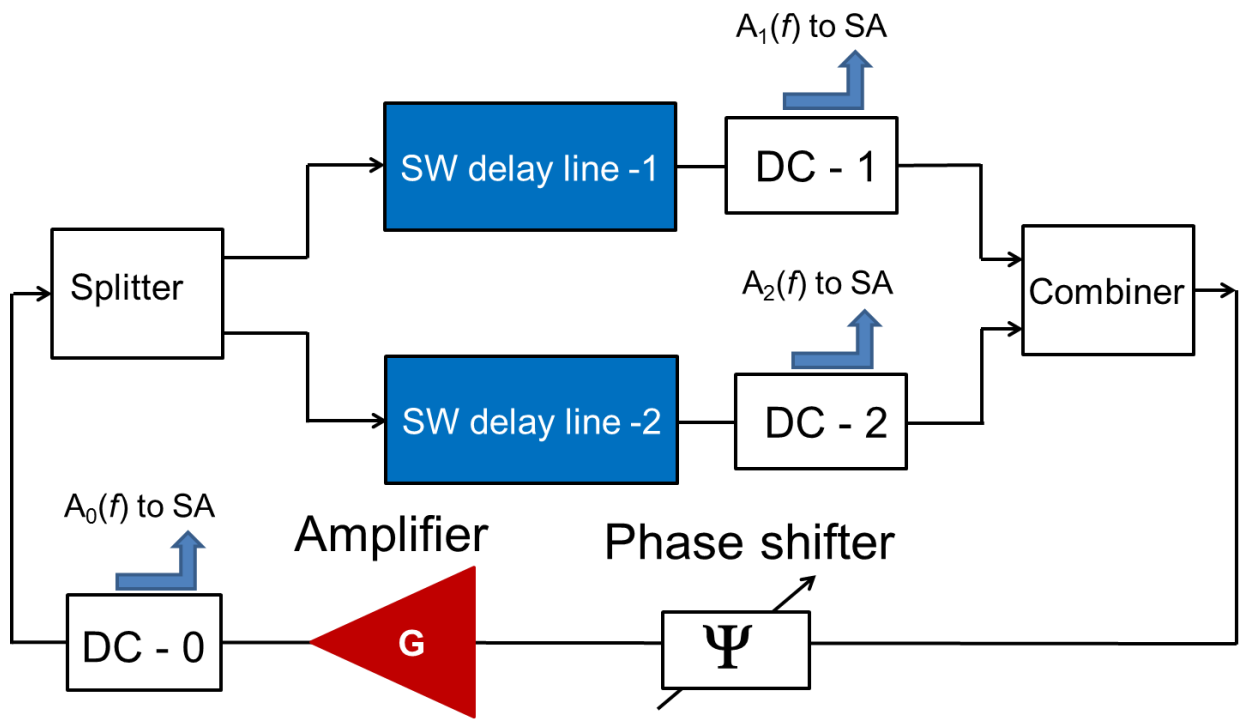

**Figure 10**

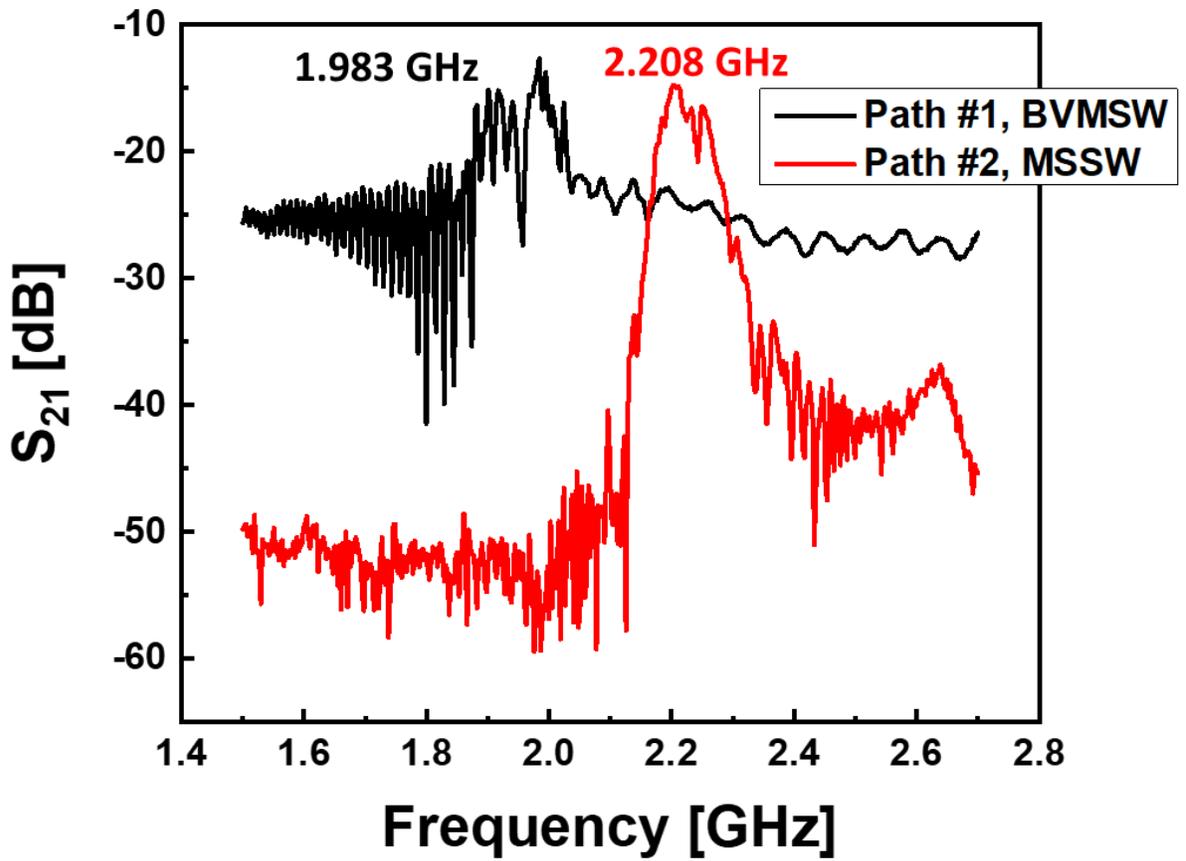

**Figure 11**

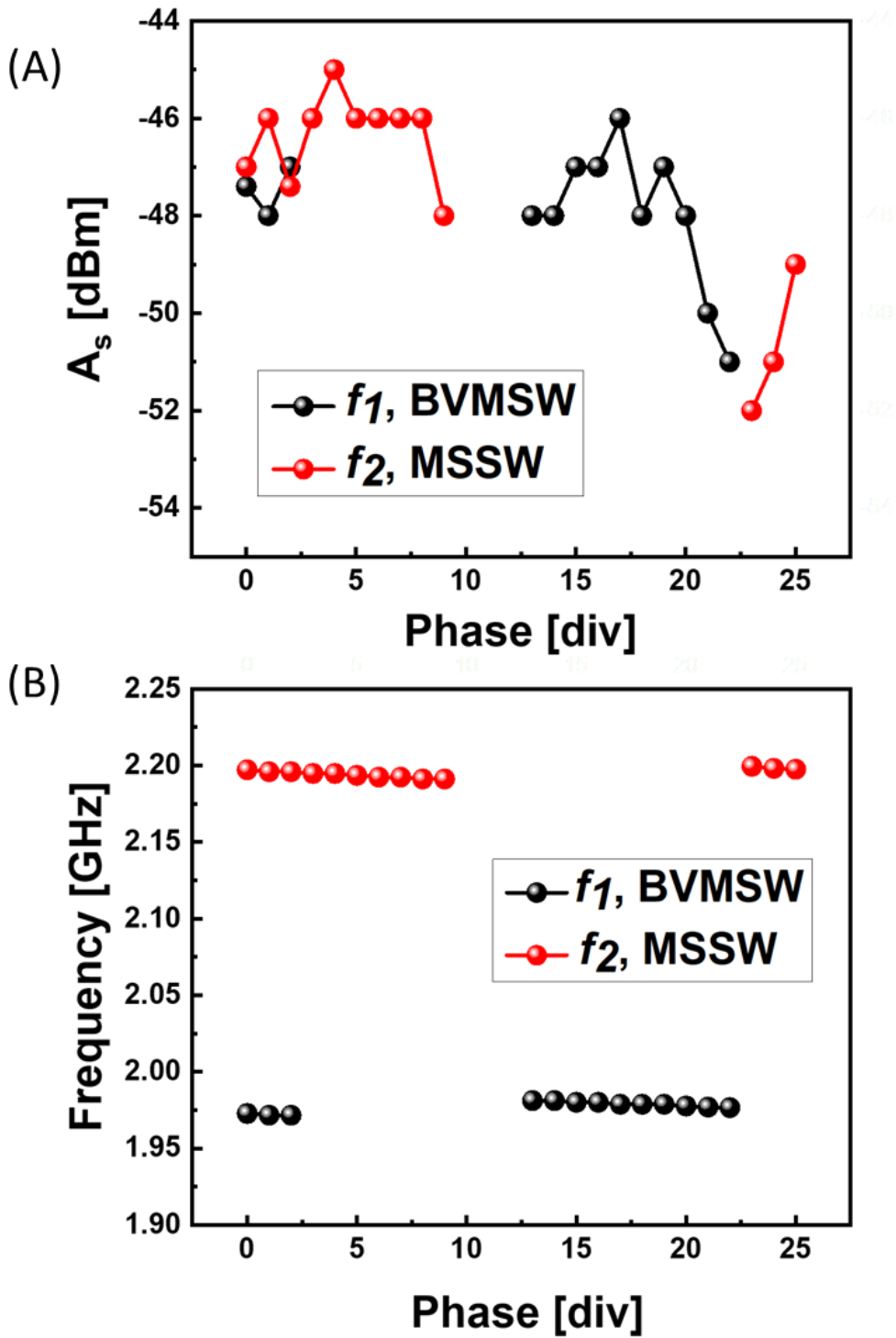

**Figure 12**

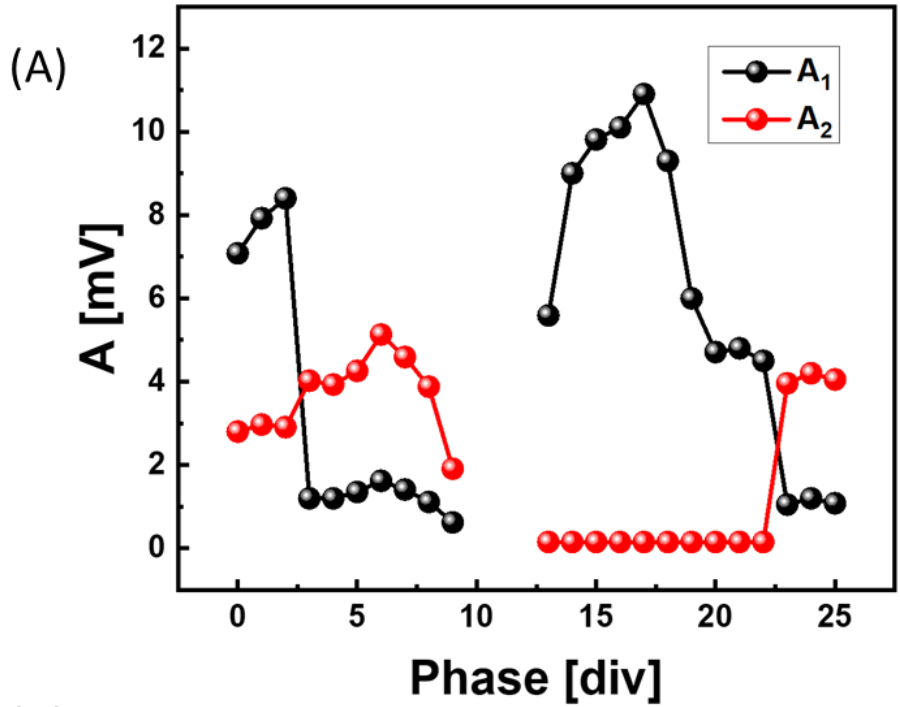
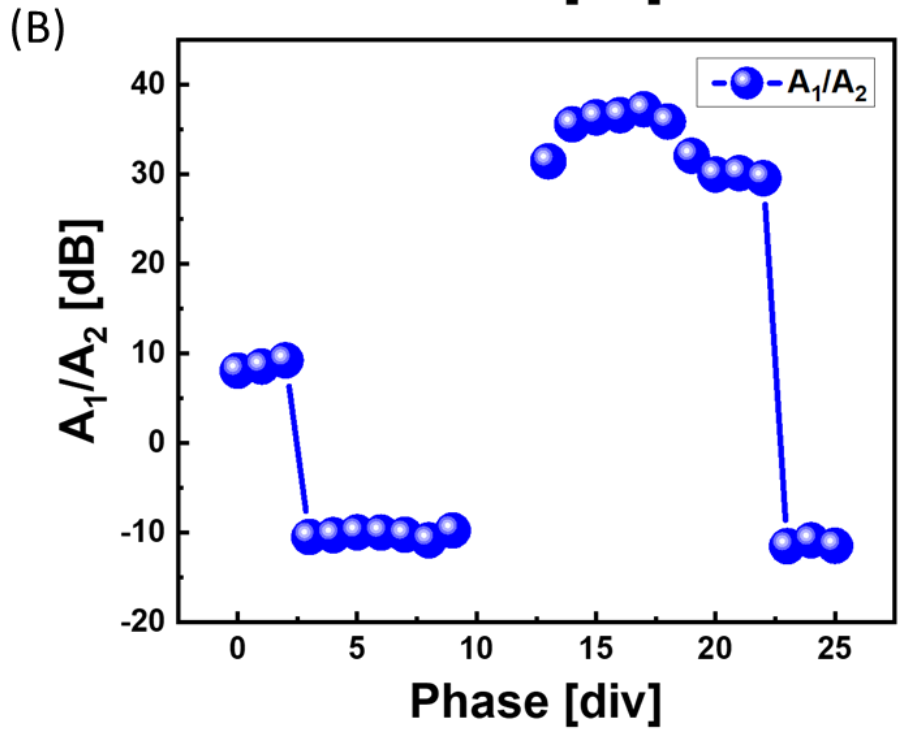

**Figure 13**

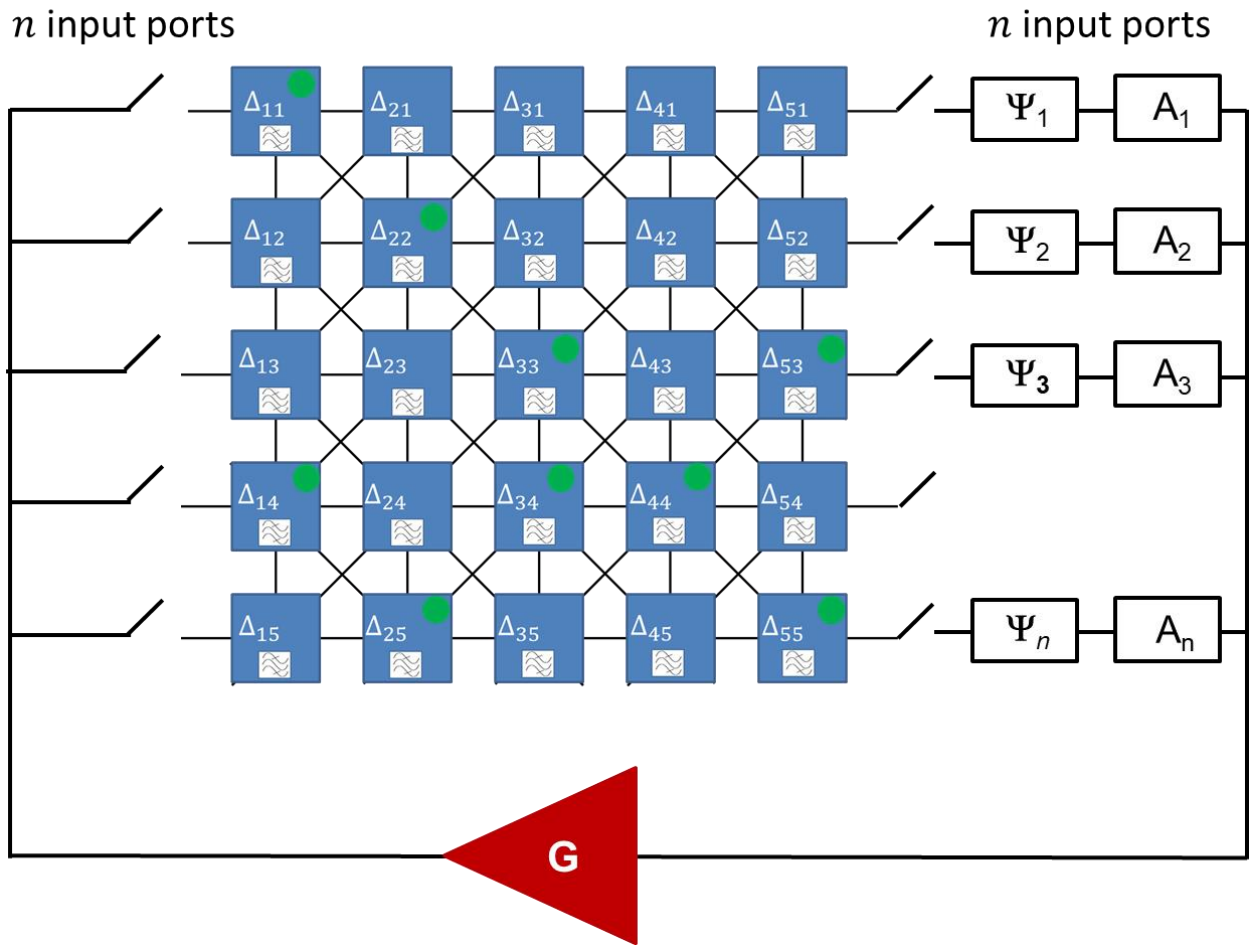

**Figure 14**